\documentclass[showpacs,twocolumn,amsmath,amssymb,pra,aps]{revtex4}

\usepackage{amssymb}
\usepackage{graphicx}
\usepackage{dcolumn}
\usepackage{bm}
\usepackage{amsmath}

\newcommand{\ket}[1]{| #1 \rangle}

\begin{document}

\title{Splitting in the excitation spectrum of a Bose-Einstein condensate undergoing strong Rabi oscillations}

\author{E. Rowen}

\author{R. Ozeri}
\author{N. Katz}
\author{ E. Gershnabel}
\author{ N. Davidson}
\affiliation{Department of Physics of Complex Systems,\\
Weizmann Institute of Science, Rehovot 76100, Israel}
\date{\today}

\begin{abstract}

We report on a measurement of splitting in the excitation spectrum
of a condensate driven by an optical travelling wave. Experimental
results are compared to a numerical solution of the Gross
Pitaevskii equation, and analyzed by a simple two level model and
by the more complete band theory, treating the driving beams as an
optical lattice. In this picture, the splitting is a manifestation
of the energy gap between neighboring bands that opens on the
boundary of the Brillouin zone.
\end{abstract}

\maketitle

Perturbative bulk excitations over  the ground state of a
weakly-interacting Bose Einstein condensate (BEC) have been
extensively studied using two-photon Bragg transitions
\cite{Stringari-structure,BlakieBallagh,ketterle-bragg,ketterle-phonon}.
Their spectrum \cite{ours-bragg}, dynamics \cite{ours-tomography}
and decay \cite{ours-beliaev} are rather well understood. However,
Bragg processes can also strongly excite the condensate,
generating non-perturbative dynamics, such as Rabi oscillations
between two (or more) macroscopically-populated momentum states
\cite{phillips-lattice,ours-dephasing,wilson-nonadiabatic}. Rapid
oscillations have been found to suppress various inhomogeneous
broadening mechanisms, thus  increasing the coherence time of the
system \cite{ours-dephasing}.

 Since the Bragg process may be
viewed as the diffraction of an atomic wavepacket from a nearly
perfect optical lattice, band theory may be employed for the study
of such systems. A BEC in an optical lattice has proved to be an
excellent system for the investigation of  effects predicted long
ago in  solid state physics such as Bloch oscillations and
Landau-Zener intra-band tunnelling
\cite{arimondo-BO,inguscio-JJ,kasevich,phillips-lattice},  and the
Mott-insulator phase transition \cite{bloch-Mott}. Much progress
has been made in understanding the ground state of a BEC in an
optical lattice, its weak excitations and the effects of
interactions \cite{Stringari-lattice,niu}. However strong
excitations involve two beating macroscopically occupied modes,
and lead to Rabi like oscillations. Thus they cannot be described
as a Bogoliubov-like excitation over a single Bloch state.

In this paper we study the spectrum of a BEC driven by a strong
resonant Bragg pulse, which leads to  coherent Rabi oscillations
between momentum states. We observe these oscillations in the time
domain and a splitting between the energy bands in the frequency
domain. The splitting is measured experimentally with Bragg
spectroscopy, using an additional Bragg beam-pair as a weak probe,
and found to agree with numerical solution of the time-dependent
Gross-Pitaevskii equation (GPE). The splitting is also predicted
by a simplified two-level dressed-state model. However some
features are explained only by the more complete band theory.
These theories do not include inhomogeneous and finite-size
effects, nor interactions, but are in reasonable agreement with
our experimental results.

The energy spectrum of a linear system is related to the time
dynamics via Fourier transform. Therefore the oscillations in the
time domain are translated to splitting in the spectrum, and the
decay of the time-correlation function results in broadening of
the  spectral peaks \cite{tannor-book}. In our experiment the
chemical potential, which characterizes the non-linearity, is
smaller than the Rabi frequency, but of the same order of
magnitude. Hence this linear description is only approximately
true for our system.

 Our experimental
apparatus is described in \cite{ours-bragg}. Briefly, a nearly
pure ($\sim 90\%$) BEC of $5 (\pm 1)\times 10^{4}$ $^{87}$Rb atoms
in the $|F,m_{f}\rangle =|2,2\rangle $ ground state, is formed in
a  magnetic trap with radial and axial trapping frequencies of
$\omega_r=2\pi \times 226$ Hz and $\omega_z=2\pi \times 26.5$ Hz,
respectively. The condensate is driven  by a pair of strong
counter-propagating Bragg beams with wave-vectors $k_L \hat{z}$
and $-k_L \hat{z}$ generating an optical lattice potential along
the axial direction  with a depth characterized by a Rabi
frequency $\Omega_d$. The laser frequency is red detuned 44 GHz
from the $^{87}$Rb D$_2$ transition in order to avoid spontaneous
emission. As shown in \cite{ours-dephasing}, in the Rabi regime
the mean field shift is largely suppressed, hence the frequency
difference between the driving Bragg beams (in the laboratory
frame) is set to $\omega_d=2\pi\times15$ kHz, the free-particle
resonance.

\begin{figure}[tb]
\includegraphics[width=8cm]{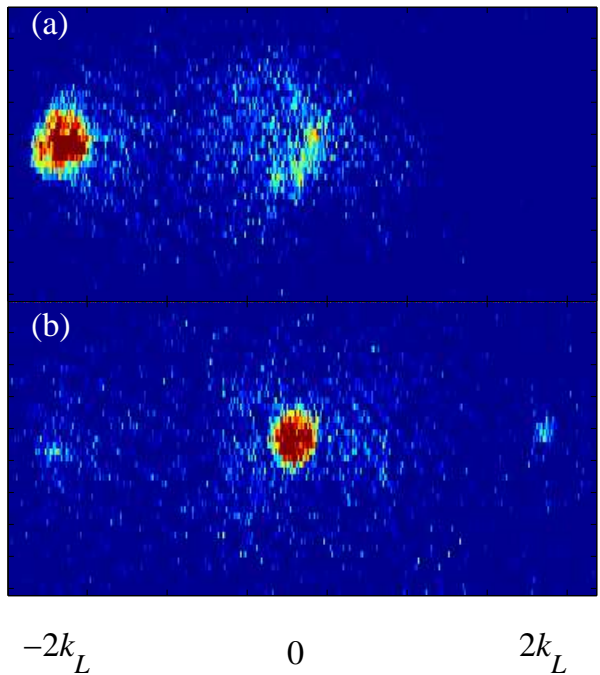}
\caption[Time of flight images]{\label{fig:tof}Time of flight
images of BEC undergoing strong Rabi oscillations. (a) BEC evolved
in the presence of driving Bragg beams alone. The pulse of
duration $320$ $\mu$sec is just over $3\pi/\Omega_d$, leaving
almost all population in $-2k_L$ momentum state. (b) Both driving
Bragg beams and probe Bragg  beams are on for $600$ $\mu$sec.
Driving pulse is close to $6\pi/\Omega_d$ leaving the state
$-2k_L$ almost unoccupied. }
\end{figure}

After driving pulses of varying duration, the magnetic trap is
rapidly turned off. A typical absorption image after 38 msec time
of flight expansion is shown in Fig. \ref{fig:tof}a. The
wavepackets with wavenumber $-2k_L$ and $k=0$ are clearly
separated. In this image, $\Omega_d t\sim 3\pi$, hence most atoms
are in the $-2k_L$ wavepacket.  The number of atoms in each
wavepacket is  measured by integration over the density in an area
determined by a gaussian fit to the absorption image. We thus
define $N_{-2k_L}\text{ and }N_{0}$ as the number of atoms in the
wavepackets with wavenumbers $-2k_L\text{ and } 0$ respectively.

\begin{figure}[tb]
\includegraphics[width=8cm]{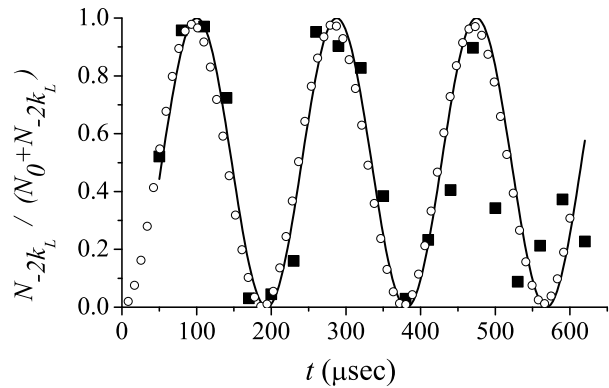}
\caption{ \label{fig:osc} Rabi oscillations between modes with
wavenumbers $0$ and $-2k_L$ (in the laboratory frame of
reference). The solid line is a sinusoidal fit to the data which
gives an oscillation frequency of $2\pi\times5.34 \text{ kHz}$.
The open circles are the GPE simulation with
$\Omega_d=2\pi\times5.5\text{ kHz}$ whose sinusoidal fit gives the
same oscillation frequency. The oscillation frequency is less than
$\Omega_d$ due to interactions, which in effect lower the depth of
the optical lattice. }
\end{figure}

Figure \ref{fig:osc} shows the time dynamics of the BEC exposed to
the driving Bragg beams. The solid squares are data points,
measuring the fraction  of atoms in $-2k_L$ cloud as a function of
the duration of the driving pulse. The BEC undergoes oscillations
at a frequency of $2\pi\times5.3\pm0.1 \text{ kHz}$ as found from
a sinusoidal fit to data. The  hollow points are numerical
solutions of the GPE including finite time, and spatial
inhomogeneity with $\Omega_d$ as the only fit parameter
\cite{ftnt:gpe}. The Rabi frequency which yields the correct
oscillation frequency is found to be $\Omega_d=2\pi\times 5.5$
kHz. The oscillation frequency is slightly less than $\Omega_d$
due to interactions. This is because the higher density in the
valleys of the optical potential  creates a mean-field lattice
potential opposing the optical one \cite{niu,arimondo-BO}.

\begin{figure}[tb]
\includegraphics[width=8cm]{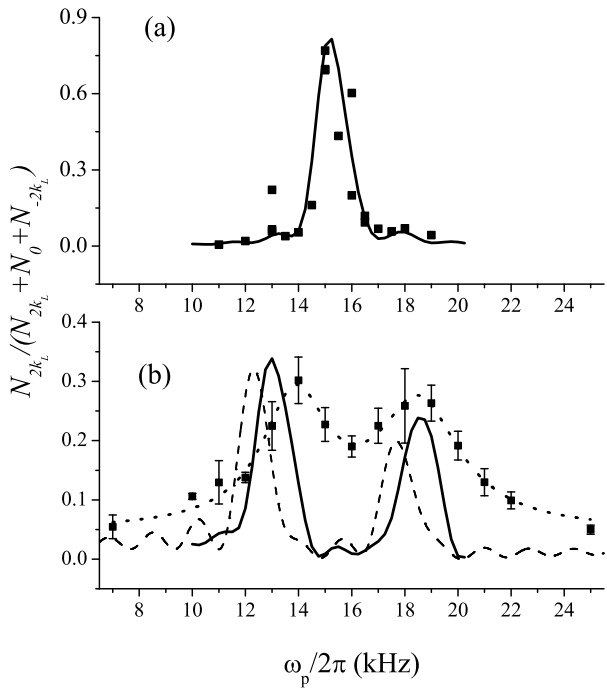}
\caption{ \label{fig:spec} Excitation spectra: Fraction of atoms
in $2k_L$ state as a function of the probe frequency difference.
(a) No lattice ($\Omega_d=0$) and (b) $\Omega_d=5.5$ kHz. In (a)
The solid line is a GPE simulation with $\Omega_p$ as a single
free parameter. In (b) the solid line is a GPE simulation with
$\Omega_p$ the same as in (a), and $\Omega_d$ determined by Fig.
\ref{fig:osc}. The dotted line is a Lorentzian fit to the data
giving a splitting of $\Delta E=h\times 4.7(\pm0.2)$ kHz. The
dashed line is the noninteracting band model discussed in the
text.}
\end{figure}

Next, we probe the spectrum of the oscillating BEC, with another
pair of Bragg beams (turned on simultaneously with the driving
Bragg beams), generating a weak lattice with a Rabi frequency
$\Omega_p$. The probe beams, also counter-propagating along the
axial direction, are linearly polarized perpendicular to the
driving beams. The frequency difference  between probe beams
$\omega_p$ is always kept such that the $0$ momentum wavepacket is
diffracted by the probe lattice to a $2k_L$ momentum wavepacket
whereas for the $-2k_L$ wavepacket, the Bragg-frequency of the
probe beams is always doppler-shifted away from resonance. For the
same reason, the driving beams are Doppler-shifted from the $2k_L$
state. A characteristic time of flight image is shown in
Fig.~\ref{fig:tof}b. Here in addition to the clouds in states
$-2k_L$ and $0$, there are also atoms with wavenumber $2k_L$. The
measured relative population of this momentum state
$N_{2k_L}/(N_{-2k_L}+N_{0}+N_{2k_L})$ as a function of the probe
frequency difference $\omega_p$ forms the probe Bragg spectrum,
which is plotted in Fig.~\ref{fig:spec} (black squares). The
duration of the beams is set so the driving pulse would be close
to an even multiple of $\pi$, in order to minimize collisions in
the time of flight expansion.

 The probe spectrum, without
the driving beams ($\Omega_{d}=0$) is similar to the well known
Bogoliubov spectrum (Fig.~\ref{fig:spec}a). The probe is weak in
the sense that $\Omega_p\simeq\Omega_d/8$, but since the probe
pulse is greater than $\pi/2$ we observe a suppression in the mean
field shift \cite{ours-dephasing}. The probe spectrum in the
presence of the driving beams (Fig. ~\ref{fig:spec}b) shows a
clear splitting in the spectrum of the oscillating condensate. The
 dotted line is a double Lorentzian fit to the experimental data, with a
  distance between the peaks of $\Delta E/h = 4.7\pm0.2 \text{ kHz}$,
  comparable but larger than the chemical potential $\mu/h=1.6$ kHz. The solid line is the full
GPE simulation of the experiment, which gives a splitting of
$\Delta E/h = 5.5 \text{ kHz}$. We note this is not the frequency
of oscillations in Fig.~\ref{fig:osc}, but rather the Rabi
frequency of the lattice itself. The mismatch between experimental
and simulation results, may be due to a drift in the intensity of
the laser beams between the oscillations measurement and the
splitting measurement.

The width of the spectrum without driving (Fig. \ref{fig:spec}a)
fits the GPE simulation, and is mostly due to finite time of the
experiment. However the experimental peaks in Fig. \ref{fig:spec}b
are substantially broader than predicted by the GPE. This
broadening is robust and is observed for different values of
$\Omega_d$, indicating broadening mechanisms which are beyond mean
field. One such mechanism for a homogeneous condensate is due to
uncertainty in a dressed state decomposition of the initial Fock
state, that leads to a spectral broadening
\cite{ours-wave_mixing}. For our experimental parameters this
 broadening is much less than the widths of
the experimental peaks which are $3\pm 0.4$ and $3.7\pm 0.5$ kHz.
We do not fully understand the broadening.  The overall response
of the system to the probe is smaller in Fig.~\ref{fig:spec}b
(with driving beams) than that in Fig.~\ref{fig:spec}a (without
driving beams). This is because the time-averaged population of
the state $k=0$ in the presence of the lattice is only half of
that of the ground state of the BEC.


The simplest model that explains the splitting in the spectrum is
a noninteracting model of a closed system with two momentum
states: $0$ and $2\hbar k_L$, coupled to a classical light
potential. Choosing our frame of reference  as moving with the
optical lattice of the Bragg driving beams at velocity $v=-k_L/M$
along z yields a time-independent Hamiltonian,

\begin{equation} H_0=-\frac{\hbar^2}{2M}\frac{\partial^2}{\partial z^2}+\hbar\Omega_d
\cos(2k_Lz) \label{eq:bloch hamiltonian}.
\end{equation}

The eigenstates of such a Hamiltonian are dressed states
\cite{API}. The energy difference between the two eigenenenergies
is $\Omega_d$. This energy difference is manifested both in the
splitting of the spectrum, and in the Rabi oscillations which are
the coherent beating between the two eigenstates. Since the
initial $k=0$ state is an equal super-position of the two dressed
states, the two peaks in the spectrum are expected to have the
same hight.

The system is however not a closed two-state system, as there are
nonresonant Rabi oscillations with  other momentum states. We
therefore calculate the eigenstates of the non-interacting
Hamiltonian~(\ref{eq:bloch hamiltonian})  without the restriction
to two momentum states. According to Bloch's theorem, due to the
periodicity of the optical lattice, a state $|k\rangle$ with
wavenumber $k$ is only coupled to states $|k+2mk_L\rangle$, where
$m$ is an integer, and $2k_L$ is the basis of the reciprocal
lattice. The Hamiltonian is diagonalized by the Bloch states
\begin{equation}
\label{eq:blochStates} |n,q\rangle=\sum_m
a_{n,q}(k)|q+2mk_L\rangle
\end{equation}
 where $q$ is the
\emph{quasi-momentum} and $n$ is the band index
\cite{BlakieBallagh}. In the moving frame of reference, the
stationary BEC has momentum $\hbar k_L$ and is situated on the
Brillouin zone boundary. The initial kinetic energy of the
condensate is in the energy gap, and by suddenly turning on the
light potential, the initial state $|k_L\rangle$ is projected onto
the Bloch states $|k_L\rangle=\sum_n a_{n}(k_L)|n\rangle$, where
we have omitted the quantum number $q$ which is understood to be
$q=k_L$.

\begin{figure}[tb]
\begin{center}
\includegraphics[width=8cm]{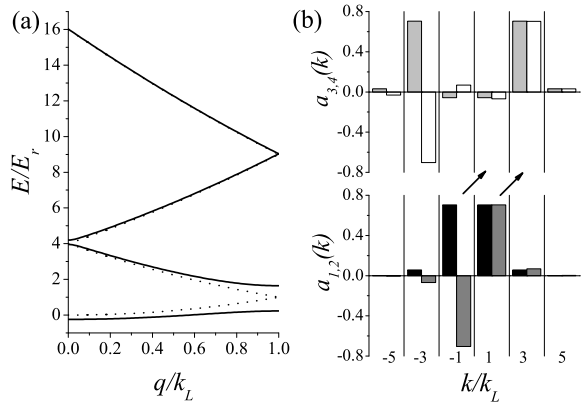}
\end{center}
\caption{ \label{fig:bands} Noninteracting band structure for our
experimental parameters. (a) The energy of the four lowest bands
as a function of quasi-momentum (solid lines). The energy gap on
the boundary of the Brillouin zone is $0.99$ $\hbar\Omega_d$,
while the gap between the third and fourth band is two orders of
magnitude smaller. The dashed line is the free particle energy.
(b) The amplitudes of the different momentum states $k=k_L+2mk_L$
in each Bloch state on the boundary of the Brillouin zone. Black,
dark gray, light gray and white are bands 1,2,3 and 4
respectively. The arrows represent the coupling by the probe. Note
the contamination of each Bloch state by the ``unintuitive''
momentum states.}
\end{figure}

In the weak lattice limit, $\hbar\Omega_d\ll E_r$ where
$E_r=\hbar^2k_L^{2}/2M$ is the single photon recoil energy, we
recover the two state result
$|k_L\rangle=1/\sqrt{2}\left(\left|1\right\rangle+\left|2\right\rangle\right)$,
and a splitting of $\Omega_d$. For our parameters the band
structure is obtained by numerical diagonalization and presented
in Fig. \ref{fig:bands}. The predicted splitting between the two
lowest bands  is indeed within $1\%$ of $\Omega_d$, but contrary
to the weak lattice approximation, $|a_{3}(\pm k_L)|=|a_{4}(\pm
k_L)|\approx 10\%$ of $|a_{3}(\pm3k_L)|=|a_{4}(\pm3k_L)|$. This
indeed leads to negligible occupations, but as we show shortly, to
observable interference effects.

With the probe beams, the Hamiltonian governing the model
homogeneous system in the driving lattice frame is $H=H_0+H_p$
with
\begin{equation} H_p=\hbar\Omega_p \cos(2k_Lz-\omega_p t)=V_pe^{-i\omega_p t}+c.c.\label{eq:probe
hamiltonian}
\end{equation}
with $V_p=\hbar\Omega_p e^{2ik_Lz}/2$. In the rotating wave
approximation $H_p$ is a momentum shift operator coupling state
$|k\rangle \text{ to } |k+2k_L\rangle$. Since
$\Omega_p<<\Omega_d$, it can be treated by the use of first-order
time-dependant perturbation theory, giving probability amplitudes
for transitions between bands. The momentum transferred to the
condensate by the redistribution of photons in
 the probe beams is a multiple of the reciprocal lattice vector
 $2k_L$, hence also with the probe beams
the initial quasi-momentum $q=k_L$ is conserved.

 The dashed line in Fig.~\ref{fig:spec}b
is the full first order perturbation theory expectation of the
population of state $|3k_L\rangle$ in the moving frame of
reference, which is  equal to $2k_L$ in the laboratory frame of reference.
 Aside from a mean-field shift, this simple model seems to fit the GPE simulation quite well.

Since the splitting between band 3 and 4 is negligible, the
splitting in the spectrum of the probe  in this non-interacting
model is essentially the splitting between the two lowest bands
(see Fig. \ref{fig:bands}a).

There is a clear asymmetry in the peak heights which is in good
agreement with the GPE simulation. This asymmetry is due to
quantum interferences between pathways of the probe excitation.
The probe couples each of the lower bands (1 and 2)  to each of
the higher bands (3 and 4) by two paths. In the weak lattice
approximation there is only the path
$\ket{k_L}\rightarrow\ket{3k_L}$, however due to the nonvanishing
$a_{3}(k_L)$ and $a_{4}(k_L)$, there is also a path
$\ket{-k_L}\rightarrow\ket{k_L}$. These paths are marked by arrows
connecting the relevant amplitudes in Fig 4(b). The interference
between both paths is constructive for transitions from band 2 and
destructive for transitions from band 1. Therefore the lower
energy peak in the splitting, corresponding to transitions from
band 2 should be larger than the higher energy peak which
corresponds to transitions from band 1.

\begin{figure}[tbh]
\begin{center}
\includegraphics[width=8cm]{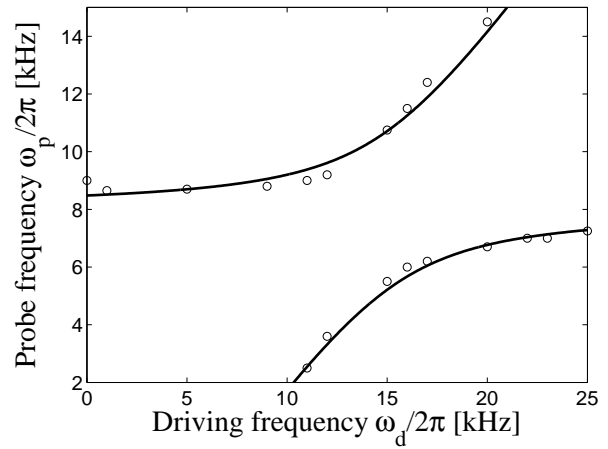}
\end{center}
\caption{ \label{fig:detunig} The locations of the peaks in the
response to the probe, as a function of the driving beams'
frequency difference $\omega_d$. (circles) - numerical solution of
the GPE including finite size, finite time and mean field effects.
(lines) - the energy levels of a two-level system shifted up by
1.3 kHz. The probe wavenumbers are taken as $\pm k_L/3$ to
distinguish probe excitations from non-resonant driving
excitations. }
\end{figure}

In the experiment, the frequency difference between the driving
Bragg beams is equal to the free particle resonance
$\hbar\omega_d=4E_R$. For other values of $\hbar\omega_d$, the
Rabi oscillations are not full. As the detuning from resonance
grows, the amplitude of oscillations is reduced and the frequency
of oscillations grows. In the spectral domain the frequency
difference between the two peaks of Fig. \ref{fig:spec}b grow with
the detuning from resonance. The amplitude of one peak grows while
it's position in frequency shifts towards the excitation energy of
a free Bogoliubov quasiparticle. The other peak shrinks in size as
the detuning of $\omega_d$ from resonance grows. In Fig.
\ref{fig:detunig} we plot the values of $\omega_p$ at the peak of
the spectrum  versus $\omega_d$, as obtained by numerically
solving the GPE \cite{ftnt:gpe} for $\Omega_d=5.5$ kHz. We compare
this to the energy splitting of a non-interacting two-level
system, calculated for the same parameters, and shifted up  by
$1.3$ kHz. As in the resonant case, the splitting is in good
agreement with the GPE, except for a constant shift due to
interactions.

In conclusion, we  measure a splitting in the Bragg excitation
spectrum of a BEC undergoing Rabi oscillations between two
momentum states. Experimental data fits well with numerical GPE
simulations. The main features are captured by a simple
noninteracting model, whereas the main contribution of
interactions is found to be  the mean field shift of the probe
spectrum, captured by the GPE, and an increase in the width which
is not captured by the GPE. The splitting in the spectrum results
in a splitting in the energy of the atoms undergoing incoherent
collisions. Experimentally, a divergence from the known s-wave
scattering halo was measured \cite{ours-collisions} for a rapidly
oscillating BEC. There too,  a pronounced asymmetry was found
between the inner collisional shell and the outer one.

This work was supported in part by the Israel Ministry of Science,
the DIP foundation, the Israel Science Foundation and by the
Minerva foundation.


\end{document}